%% file: prl.tex
\RequirePackage[switch]{lineno}
\documentclass[twocolumn,showpacs,aps,prd,superscriptaddress]{revtex4}
\usepackage{graphicx}
\usepackage{dcolumn}
\usepackage{amsmath}
\usepackage{epsfig}
\RequirePackage{xspace}

\usepackage{relsize}
\def\babar{\mbox{\slshape B\kern-0.1em{\smaller A}\kern-0.1em
    B\kern-0.1em{\smaller A\kern-0.2em R}}}

\def\lsim{{~\raise.15em\hbox{$<$}\kern-.85em
          \lower.35em\hbox{$\sim$}~}\xspace}

\def\en         {\ensuremath{e^-}\xspace}
\def\ep         {\ensuremath{e^+}\xspace}

\def\epem       {\ensuremath{e^+e^-}\xspace}

\def\mup        {\ensuremath{\mu^+}\xspace}
\def\mun        {\ensuremath{\mu^-}\xspace}
\def\mumu       {\ensuremath{\mu^+\mu^-}\xspace}

\def\taup       {\ensuremath{\tau^+}\xspace}
\def\taum       {\ensuremath{\tau^-}\xspace}

\def\ellell     {\ensuremath{\ell^+ \ell^-}\xspace}
\def\g     {\ensuremath{\gamma}\xspace}
\def\gaga  {\ensuremath{\gamma\gamma}\xspace}
\def\ggstar{\ensuremath{\gamma\gamma^*}\xspace}

\def\pep2{PEP-II}

\def\qqbar {\ensuremath{q\overline q}\xspace}

\def\dbar  {\ensuremath{\overline d}\xspace}

\def\bbar  {\ensuremath{\overline b}\xspace}

\newcommand{\CP}{$CP$}
\newcommand{\gevcc}{~{\rm GeV}/c^2}
\newcommand{\gev}{~\rm GeV}
\newcommand{\gevc}{~{\rm GeV}/c}
\newcommand{\mev}{~\rm MeV}
\newcommand{\mevcc}{~{\rm MeV}/$c^2$}
\newcommand{\etal}{$\it et~al.$}
\newcommand{\FourS}{\Upsilon (4S)}
\newcommand{\kskspi}{\KS\KS \pip}
\newcommand{\btokskspi}{\Bp \to \KS \KS \pip}
\newcommand{\btoksksk}{\Bp \to \KS \KS \Kp}
\newcommand{\btodpi}{\Bp \to {\Dzb}(\to \KS\KS) \pip}
\newcommand{\invfb} {~$\rm fb ^{-1}$}
\newcommand{\onreslumi}  {\mbox{423.7\invfb}}
\newcommand{\offreslumi}  {\mbox{43.9\invfb}}
\newcommand{\bbpairs}    {\mbox{$(465.1\pm 5.1)\times10^{6}$}}
\def\B       {\ensuremath{B}\xspace}
\def\Bbar    {\kern 0.18em\overline{\kern -0.18em B}{}\xspace}
\def\Dbar    {\kern 0.18em\overline{\kern -0.18em D}{}\xspace}

\def\BB      {\ensuremath{B\Bbar}\xspace}
\def\Bz      {\ensuremath{B^0}\xspace}
\def\Bzb     {\ensuremath{\Bbar^0}\xspace}
\def\Dzb     {\ensuremath{\Dbar^0}\xspace}
\def\BzBzb   {\ensuremath{\Bz {\kern -0.16em \Bzb}}\xspace}
\def\Bu      {\ensuremath{B^+}\xspace}
\def\Bub     {\ensuremath{B^-}\xspace}
\def\Bp      {\ensuremath{\Bu}\xspace}

\def\Kz    {\ensuremath{K^0}\xspace}
\def\Kzb   {\ensuremath{\Kbar^0}\xspace}
\def\kaon  {\ensuremath{K}\xspace}
\def\Kbar  {\kern 0.2em\overline{\kern -0.2em K}{}\xspace}

\def\KzKzb {\ensuremath{\Kz \kern -0.16em \Kzb}\xspace}
\newcommand{\Kp}{K^+}
\newcommand{\Km}{K^-}
\def\KS    {\ensuremath{K^0_{\scriptscriptstyle S}}\xspace}
\def\KL    {\ensuremath{K^0_{\scriptscriptstyle L}}\xspace}
\newcommand{\pip}{\pi ^+}
\newcommand{\pim}{\pi ^-}
\newcommand{\piz}{\pi ^0}
\newcommand{\mes}{m_{\rm ES}}
\newcommand{\DeltaE}{\Delta E}
\newcommand{\bdt}{\rm BDT_{out}}
\def\Kstarz  {\ensuremath{K^{*0}}\xspace}

\newcommand{\BABARPubYear}    {08}
\newcommand{\BABARPubNumber}  {046}
\newcommand{\SLACPubNumber} {13461}

\def\figurebox#1#2#3{
    \def\arg{#3}
    \ifx\arg\empty
    {\hfill\vbox{\hsize#2\hrule\hbox to #2{\vrule\hfill\vbox to #1{\hsize#2\vfill}\vrule}\hrule}\hfill}
    \else
    {\hfill\epsfbox{#3}\hfill}
    \fi}

\begin{document}

\preprint{\babar-PUB-\BABARPubYear/\BABARPubNumber}
\preprint{SLAC-PUB-\SLACPubNumber}

\begin{flushleft}
\babar-PUB-\BABARPubYear/\BABARPubNumber\\
SLAC-PUB-\SLACPubNumber\\
\end{flushleft}

\title{\large \bf \boldmath Search for the decay $\Bp\to\KS\KS\pip$}

\input{pubboard/authors_sep2008.tex}

\date{\today}%
\begin{abstract}
We search for charmless decays of charged $B$ mesons to the three-body final state~$\kskspi$. Using a data sample of \onreslumi\ collected at the $\FourS$ resonance with the \babar\ detector, corresponding to \bbpairs\ \BB\ pairs, we find no significant signal and determine a 90\,\% confidence level upper limit on the branching fraction of $5.1 \times 10^{-7}$.
\end{abstract}
\maketitle

Charmless decays of $B$ mesons to final states with even numbers of strange
quarks or antiquarks, such as $\btokskspi$~\cite{cc}, are suppressed
in the standard model. Such decays proceed mainly via the $\bbar\to\dbar$
loop (penguin) transition. Hadronic $\bbar\to\dbar$
penguin transitions have been observed in the decays
$\Bz\to\Kz\Kzb$ and $\Bu\to\Kzb\Kp$~\cite{babar:ksks,belle:ksks}, 
and their effects have also been seen
through direct \CP\ violation in charmless \B decays,
such as $\Bz\to\pip\pim$~\cite{Abe:2006cc,Aubert:2007mj} and
$\Bz\to\pip\pim\piz$~\cite{Kusaka:2007dv,Aubert:2007jn}.
In contrast to $\Bz$-$\Bzb$ mixing, which is a $\bbar\to\dbar$ process 
with a change of beauty-flavor quantum number of $\Delta F=2$,
little experimental information exists on $\Delta F=1$ $\bbar\to\dbar$
decay amplitudes.
There is still potential for new physics effects to
be uncovered in these decays.

The decay $\Bu\to\KS\KS\pip$ has not yet been observed.
The upper limit on the branching fraction at 90\,\% confidence level (CL) is 
$3.2\times 10^{-6}$~\cite{belle:kskspi}.
A model based on the factorization approximation, which makes use of 
heavy-quark and chiral symmetries, predicts a nonresonant branching 
fraction for $\Bp\to\Kz\Kzb\pip$ of order $10^{-6}$~\cite{Fajfer:1998yc}. 
Decays via intermediate resonant states can also lead to the $\KS\KS\pip$ final
state. This motivates an inclusive analysis incorporating both nonresonant
and resonant modes. Based on the measured branching fraction 
${\cal B}[\Bu\to f_2(1270)\pip]=(8.2\pm 2.5)\times 10^{-6}$~\cite{Barberio:2008fa,Amsler:2008zz,Aubert:2005sk},
the product branching fraction for $\Bu\to f_2(1270)\pip$ with 
$f_2(1270)\to\KS\KS$ should be around $10^{-7}$. 
Similarly, $\Bu\to f_0(980)\pip$ and $\Bu\to K^{*+}(892)\Kzb$ decays could
contribute to the $\KS\KS\pip$ channel. 
The branching fraction for $\Bu\to K^{*+}(892)\Kzb$ is predicted to be of order
$10^{-6}$ or less~\cite{Du:1995ff,Ali:1998eb,Du:2002up,Beneke:2003zv,Chiang:2003pm,Guo:2006uq}.

Another motivation comes from the recent observation of $\Bu\to\Kp\Km\pip$ by
\babar, with an inclusive branching fraction of
${\cal B}(\Bu\to\Kp\Km\pip)=[5.0\pm 0.5{\mathrm{(stat.)}} \pm 0.5{\mathrm{(syst.)}}]\times 10^{-6}$~\cite{babar:kkpi}.
An unexpected peak seen
near 1.5$\gevcc$ in the $\Kp\Km$ invariant-mass spectrum,
which we dub the $f_{\rm X}(1500)$,
accounts for approximately half of the total event rate.
If the decay of the $f_{\rm X}(1500)$ follows isospin symmetry, then equal 
rates would be expected to $\Kp\Km$ and to $\Kz\Kzb$.
If the $f_{\rm X}(1500)$ has even spin, then $f_{\rm X}(1500)\to\Kz\Kzb$
decays would result in 50\,\% $\KS\KS$ and 50\,\% $\KL\KL$ final states, whereas
if the $f_{\rm X}(1500)$ has odd spin, then the $\KS\KS$ final state 
is forbidden by Bose symmetry. 
Observation of the decay $f_{\rm X}(1500)\to\KS\KS$ in $\Bu\to\KS\KS\pip$
could therefore provide information on the spin or the quark content
of the $f_{\rm X}(1500)$ and could help to elucidate the relationship between
this state and similar unexplained structures seen in 
$\Bu\to\Kp\Km\Kp$ decays~\cite{belle:kkk,babar:kkk}. 
Structures in the $\KS\KS$ mass spectrum have also been observed in 
two-photon~\cite{L3:ksks} and electron-proton collisions~\cite{zeus:ksks}.

We report a search for the decay $\btokskspi$.
The analysis is based on data collected
at the \pep2\ asymmetric-energy \epem collider~\cite{pep2} at SLAC.
The data sample consists of an integrated luminosity of \onreslumi\
recorded at the $\FourS$ resonance (on-peak)
and \offreslumi\ collected 40\,\mev\ below the resonance (off-peak).
The on-peak data sample contains \bbpairs\ \BB\ pairs~\cite{Aubert:2002hc}.

The \babar\ detector is described in detail elsewhere~\cite{Aubert:2001tu}.
Charged particles are detected and their momenta measured
with a five-layer silicon vertex tracker (SVT) and a 40-layer
drift chamber (DCH) located inside a 1.5\,T solenoidal magnet. Surrounding
the DCH is a detector of internally reflected Cherenkov radiation
(DIRC), designed for charged particle identification.
Energy deposited by electrons and photons
is measured by a CsI(Tl) crystal electromagnetic calorimeter (EMC).
Muons and long-lived neutral hadrons are identified in the flux return
of the solenoid instrumented with resistive plate chambers and limited
streamer tubes.

We reconstruct a $\btokskspi$ candidate by combining a pair of $\KS$ mesons and a charged pion.
A $\KS\to\pip\pim$ candidate is formed from a pair of oppositely charged tracks
with an invariant mass that lies within 15\mevcc\  
of the nominal $\KS$ mass~\cite{Amsler:2008zz},
which corresponds to five times the $\KS$ mass resolution.
We require the ratio of measured $\KS$ lifetime and its uncertainty to be greater than 20, the
cosine of the angle between the line connecting the $B$ and $\KS$ decay vertices
and the $\KS$ momentum vector to be greater than 0.999, 
and the $\KS$ vertex probability to be greater than $10^{-6}$. Charged pions
coming from the $B$ decay are identified with 
the energy loss (d$E$/d$x$) information from the SVT and DCH, and the Cherenkov angle 
and the number of photons measured by the DIRC. The efficiency for pion 
selection is approximately 76\,\% including geometrical acceptance, while the
probability for misidentification of kaons as pions is less than 15\,\%, up to a momentum of 4$\gevc$. 
We require pion candidates not to be consistent with the electron hypothesis, based on 
information from the d$E$/d$x$, the shower shape in the EMC, and the ratio of the shower energy and track momentum.

Continuum $e^+e^-\to\qqbar~(q=u,d,s,c)$ events are the dominant background.
To discriminate this type of event from signal we use a boosted decision tree
(BDT)~\cite{ref:TMVA} that combines five discriminating variables.
The first of these is the ratio of $L_2$ to $L_0$, with $L_j = \Sigma _{i} p_i
|\cos \theta _{i}|^{j}$, where $\theta _i$ is the angle, with respect to the 
$B$ thrust axis, of the track or neutral cluster $i$, and $p_i$ is its momentum.
The sum excludes the daughters of the $B$ candidate and
all quantities are calculated in the $e^+e^-$ center-of-mass (CM) frame. The other four
variables are the absolute value of the cosine of the angle between the \B
direction and the beam ($z$) axis,
the magnitude of the cosine of the angle between the \B\ thrust axis
and the $z$ axis, the product of the \B\ candidate's charge and the flavor of the recoiling \B
as reported by a multivariate tagging algorithm~\cite{Aubert:2002ic},
and the proper time difference between the decays of the two \B\ mesons divided by its uncertainty.
The BDT is trained using off-peak data as well as simulated signal events
that pass the selection criteria. We make a requirement on the
BDT output ($\bdt$) such that approximately 96\,\% of the signal is retained
and 60\,\% of the continuum background is rejected.

In addition to $\bdt$, we distinguish signal from background
events using two kinematic variables: the beam-energy-substituted mass
$\mes = \sqrt{s/4-{\bf p}^2_\B}$ and $\DeltaE = E_\B - \sqrt{s}/2$,
where $\sqrt{s}$ is the total \epem CM energy and $(E_\B,{\bf p}_\B)$ is
the four-momentum of the \B\ candidate measured in the CM frame. We select
signal candidates that satisfy $5.250 < \mes < 5.286 \gevcc$ and 
$|\DeltaE| < 0.1 \gev$. This region includes a sufficiently large range of 
$\mes$ below the signal peak to allow properties of the continuum distribution
to be determined in the maximum likelihood fit.

Another source of background arises from $\btodpi$ decays,
where the final state particles are identical to the signal.
We reduce this background by rejecting any event containing a signal candidate
with a $\KS\KS$ invariant mass in the range $ 1.82 < M_{\KS\KS} < 1.90 \gevcc$. 

The efficiency for signal events to pass the selection criteria is 28\,\%,
determined with a Monte Carlo (MC) simulation in which decays are generated 
uniformly in three-body phase space.
We find that approximately 9\,\% of the selected $\btokskspi$ events contain
more than one candidate, in which case we choose that with the highest
 $B$-vertex probability. We have checked that this procedure does not bias the fit variables. In about 2\,\% of the signal events, the
\B candidate is misreconstructed because one of its daughter tracks is replaced by a track from the rest of the event. 
Such events are considered to be a part of the signal component.

We study possible residual backgrounds from \BB\ events using MC event samples.
These backgrounds arise from decays with similar kinematic properties to 
the signal or because particles get lost to, or attached from, 
the rest of the event in the process of reconstruction. 
The $\BB$ background modes can conveniently be divided into two categories, based
on their shapes in $\mes$ and $\DeltaE$. The first category ($\BB_1$) contains 
only $\btoksksk$ decays, which peak in $\mes$ around the $B$ mass and in 
$\DeltaE$ near~$-0.06\gev$. 
The second category ($\BB_2$) contains the remaining \BB\ backgrounds 
and is mainly combinatorial.

We perform an unbinned extended maximum likelihood fit to the
candidate events using three input variables: $\mes$, $\DeltaE$, and 
$\bdt$. For each category $j$
(signal, continuum background, $\BB_1$, or $\BB_2$), we define a probability
density function ${\cal P}_j$ (PDF), and evaluate it for each event $i$:

\begin{linenomath}
\begin{linenomath}
\begin{equation}
  \label{PDF-exp}
  {\cal P}^i_j \equiv
  {\cal P}_j(\mes^i,\DeltaE^i)\cdot {\cal P}_j(\bdt^{\it i}).
\end{equation}
\end{linenomath}
\end{linenomath}
The signal, continuum background, and $\BB_2$ background exhibit negligible
correlations between $\mes$ and $\DeltaE$, and so the PDF is further
factorized:
\begin{linenomath}
\begin{linenomath}
\begin{equation}
  \label{PDF-mesde}
  {\cal P}_j(\mes^i,\DeltaE^i) = {\cal P}_j(\mes^i) \cdot {\cal P}_j(\DeltaE^i) \, .
\end{equation}
\end{linenomath}
\end{linenomath}
The extended likelihood function is
\begin{linenomath}
\begin{linenomath}
\begin{equation}
  \label{eq:extML-Eq}
  {\cal L} =
  \prod_{k} e^{-n_k}
  \prod_{i}\left[ \sum_{j}n_j{\cal P}^i_j \right],
\end{equation}
\end{linenomath}
\end{linenomath}
where $n_j$($n_{k}$) is the yield for event category $j$($k$).

The signal $\mes$ distribution is parameterized with the sum
of a Gaussian and a Crystal Ball function~\cite{crystalBall}
while the $\DeltaE$ distribution is parameterized with a
modified Gaussian function with different widths on each side, as well as with additional
tails that can be different on each side. We fix the shape parameters to the
values obtained from the $\btokskspi$ phase-space MC sample.
The continuum background $\mes$ shape is described by an empirical
threshold ARGUS function,
$x\sqrt{1-x^2}\exp\left[-\xi(1-x^2)\right]$, with $x\equiv 2\mes/\sqrt{s}$
and $\xi$ a free parameter~\cite{Albrecht:1990am},
while the continuum $\DeltaE$ shape is modeled with a linear function.
We describe the $\mes$ and $\DeltaE$ shapes for the $\BB_1$ sample with a
two-dimensional histogram determined from MC events, which accounts for
correlations between these variables. One-dimensional histograms are used to
describe the $\mes$ and $\DeltaE$ distributions for the $\BB_2$ sample. The
$\bdt$ distributions for all components are described by one-dimensional
histograms. 
These are obtained from MC events for signal and the \BB\ background
categories. The continuum background $\bdt$ shape is determined from a
combination of off-peak data and on-peak data in a continuum-dominated
sideband of $\mes$, independent of the signal region, 
from which the expected \BB\ backgrounds have been subtracted.

The free parameters of our fit are the yields of the signal, continuum, 
and two \BB\ background categories, 
together with the $\xi$ parameter of the continuum $\mes$ shape
and the slope of the continuum $\DeltaE$ shape.

We test the fitting procedure by applying it to ensembles of simulated
experiments where events are generated from the PDF shapes as described above
for all four categories of events. We repeat the exercise with $\qqbar$ events
generated from the PDF while signal events are randomly extracted from the MC
samples. The \BB\ background events are either generated from PDF shapes or
drawn from MC samples. In all cases, these tests confirm that our fit performs
as expected. No bias is found for the value of the signal yield
observed in the data. 

The fit to $16\,739$ candidate events gives a signal yield of $15 \pm 15$
events, where the error is statistical only. The fit returns yields for the
continuum, $\BB_1$ and $\BB_2$ background categories of $15\,500
\pm 140$, $89 \pm 25$ and $1\,140 \pm 70$ events, respectively. These are
somewhat larger than the expected values for the first and last categories
and smaller for the second, a pattern that can be explained by the
correlations between these yields.

\begin{figure*}[!htb]
\center
\includegraphics[width=.65\columnwidth]{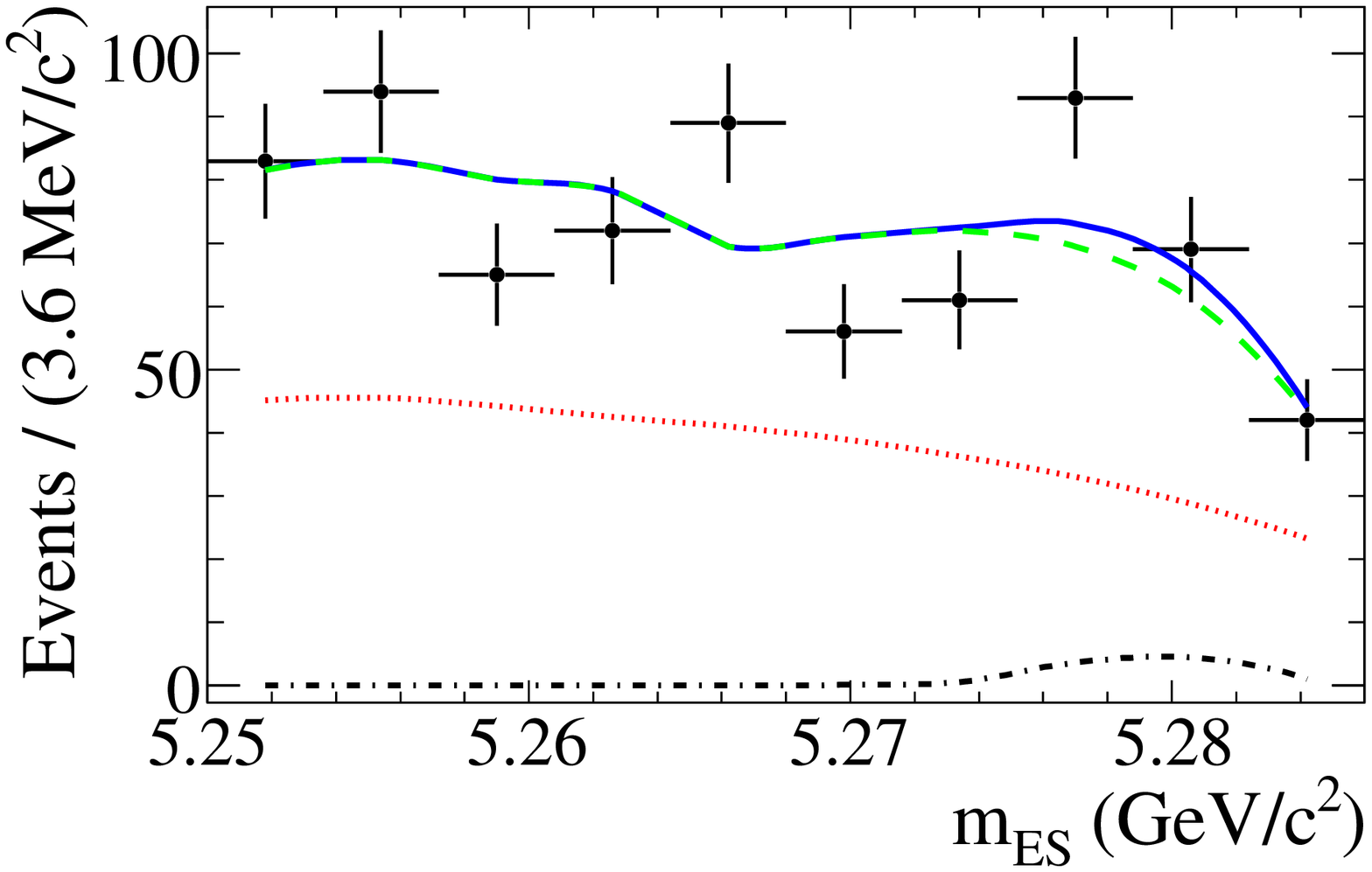}
\includegraphics[width=.65\columnwidth]{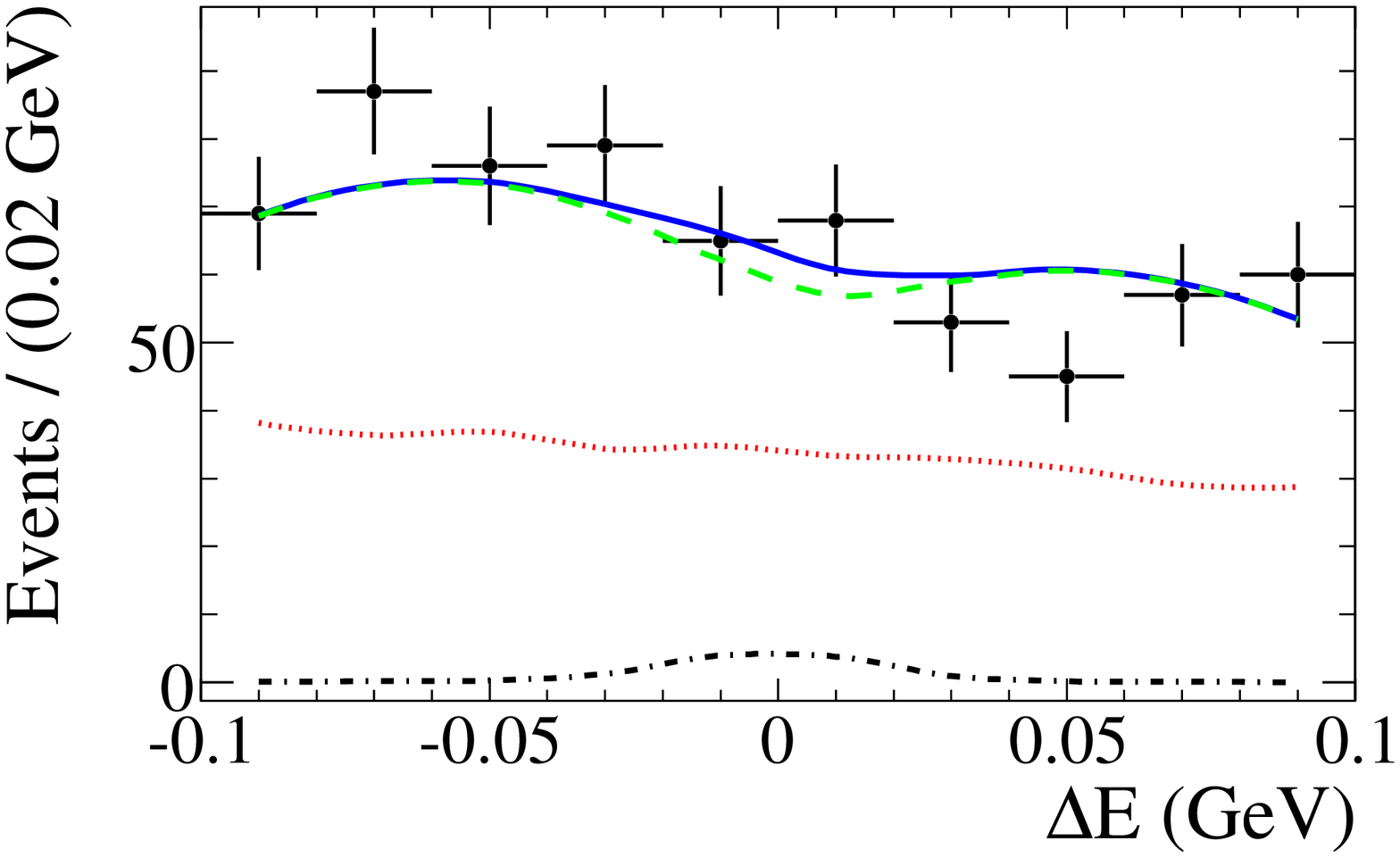}
\includegraphics[width=.65\columnwidth]{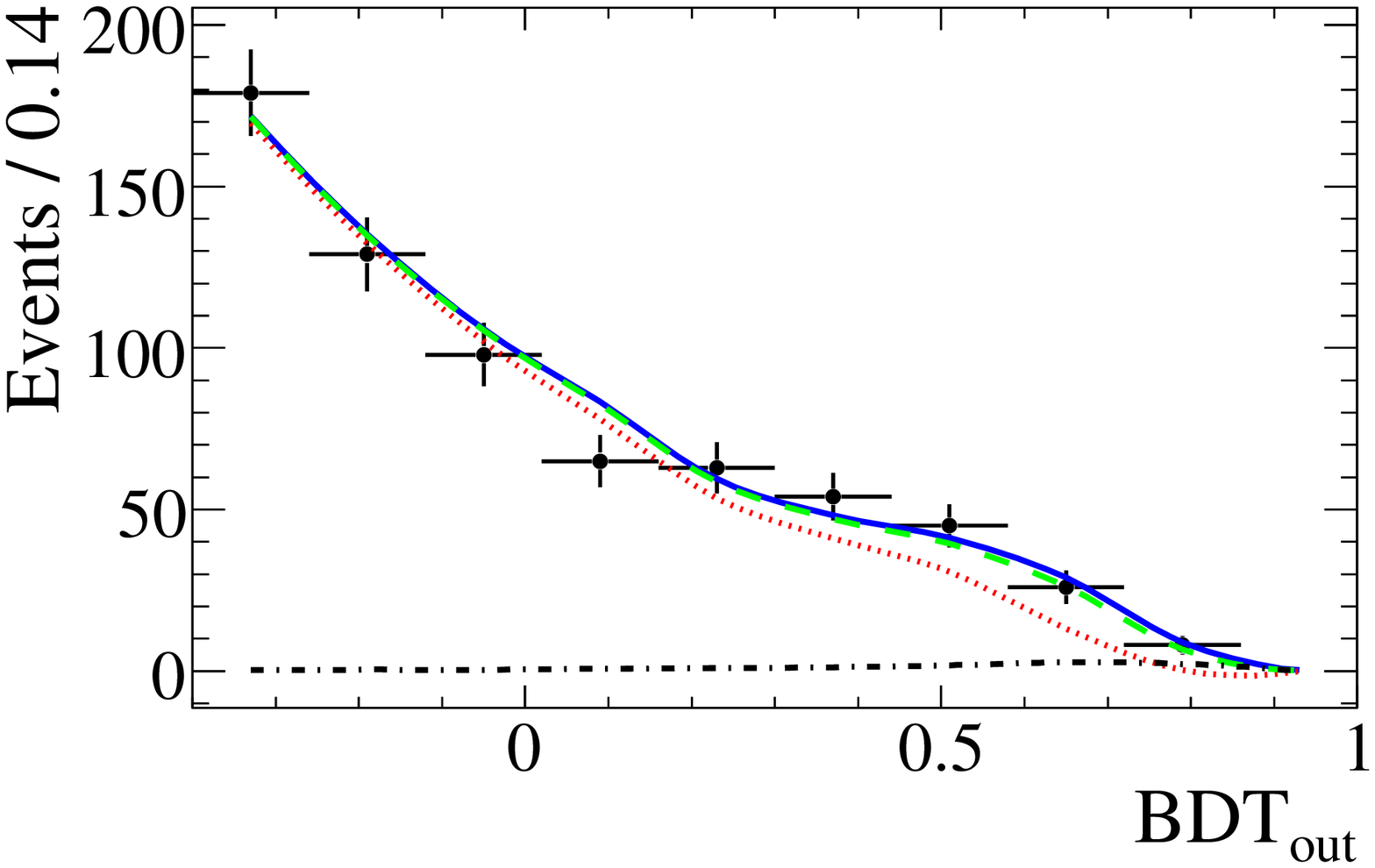}
\caption{Projections of candidate events with the fit results overlaid. 
From left to right are shown the projections onto the $\mes$, $\DeltaE$, and
$\bdt$ variables. The points show the data and the solid (blue) curves show the total fit result. The dotted (red) curves show the continuum background, the dashed (green)
curves the total background, and the dash-dotted (black) curves the signal distributions.}
\label{fig:plots}
\end{figure*}

The results of the fit are shown in Fig.~\ref{fig:plots}. In these plots
the continuum background contribution has been suppressed by applying a
requirement on the ratio of the signal likelihood to the  sum of the
signal and continuum likelihoods, calculated without use of the plotted 
variable. The value of this requirement for each plot rejects about
97\,\% of the continuum background while retaining 63 - 71\,\% of the signal,
depending on the variable.

We determine the inclusive branching fraction for $\btokskspi$
by dividing the observed signal yield by the reconstruction 
efficiency, the number of \BB\ events in the data sample,
and the square of the daughter branching fraction 
${\cal B} (\KS\ \to \pip \pim) = 0.6920 \pm 0.0005$~\cite{Amsler:2008zz}. 
We assume equal decay rates of $\FourS$ into $B^+ B^-$ and $\Bz \Bzb$ pairs. 
The value obtained is ${\cal B}(\btokskspi) = ( 2.5 \pm 2.4 ) \times 10^{-7}$, 
where the error is statistical only.
The statistical significance of the signal is $1.1~ \sigma$,
which is calculated as $\sqrt {-2\ln ({\cal L}_0/{\cal L}_{\rm max})}$, 
where ${\cal L}_{\rm max}$ denotes the likelihood with the nominal signal yield 
of 15 events and ${\cal L}_0$ denotes the likelihood with the signal yield fixed at zero.

There is a significant dependence of the selection efficiency on the
kinematics of the $\KS\KS\pip$ final state. The nominal efficiency is 
calculated by assuming a phase-space distribution of $\KS\KS\pip$ events.
Since we do not know the true distribution, a systematic uncertainty of 24\,\% 
is evaluated from the RMS variation of the efficiency across the
$\KS\KS\pip$ Dalitz plot. 
Smaller systematic uncertainties on the fitted yield arise from uncertainties 
in the PDF shapes (4 events), including possible differences between data 
and MC simulations, which are studied using a control sample of $\Bz\to D^-(\to\KS\pim)\pip$ 
events. We assign an uncertainty of 2 events to account for fit bias.
Other uncertainties on the efficiency arise from 
charged particle reconstruction (0.4\,\%), particle identification (1.4\,\%), and the $\KS$ selection (1.8\,\%).
The uncertainty on the number of \BB\ pairs is $1.1\,\%$.
The systematic uncertainties are added in quadrature
to give a total of 38\,\%.
Hence the inclusive branching fraction is ${\cal B}(\btokskspi)
=(2.5\pm 2.4\pm 0.9)\times 10^{-7}$, where the first (second) error is 
statistical (systematic).

Since our result is consistent with no signal, we determine a 90\,\% CL upper
limit on the branching fraction (${\cal B}_{\rm UL}$). This limit is calculated by
integrating the likelihood in the physical region such that 
$\int_0^{\cal B_{\rm UL}}{\cal L}(x)\, dx / \int_{0}^{+\infty}{\cal L}(x)\, dx =0.9$,
where ${\cal L}(x)$ is the likelihood function for the signal yield~$x$.
We have confirmed that the statistical uncertainties from the fit are Gaussian,
to a good approximation. We therefore assume a Gaussian behavior for the 
overall likelihood, with a width calculated
from the sum in quadrature of the statistical and systematic uncertainties.
Our result is ${\cal B}(\btokskspi) < 5.1 \times 10^{-7}$ at 90\,\% CL.

The lack of signal in this decay mode contrasts with that observed
for $\Bu\to\Kp\Km\pip$~\cite{babar:kkpi}. This result disfavors models
in which the $f_{\rm X}(1500)$ has even spin 
and decays with isospin symmetry.
If the $f_{\rm X}(1500)$ is confirmed to have even spin in future
measurements, this may indicate a non-$q\bar{q}$ nature of this state.

In conclusion, with a data sample of \onreslumi, 
we have performed a search for the decay $\btokskspi$. 
We observe no significant signal and set a 90\,\% confidence 
level upper limit on the branching fraction of $5.1 \times 10^{-7}$.
This result provides useful information for the understanding of low energy
spectroscopy.

\input{pubboard/acknow_PRL.tex}

\end{document}

%% file: pubboard/authors_sep2008.tex
%
\author{B.~Aubert}
\author{M.~Bona}
\author{Y.~Karyotakis}
\author{J.~P.~Lees}
\author{V.~Poireau}
\author{E.~Prencipe}
\author{X.~Prudent}
\author{V.~Tisserand}
\affiliation{Laboratoire de Physique des Particules, IN2P3/CNRS et Universit\'e de Savoie, F-74941 Annecy-Le-Vieux, France }
\author{J.~Garra~Tico}
\author{E.~Grauges}
\affiliation{Universitat de Barcelona, Facultat de Fisica, Departament ECM, E-08028 Barcelona, Spain }
% ...
\collaboration{The \babar\ Collaboration}
\noaffiliation

%% file: pubboard/acknow_PRL.tex
We are grateful for the excellent luminosity and machine conditions
provided by our \pep2\ colleagues, 
and for the substantial dedicated effort from
the computing organizations that support \babar.
The collaborating institutions wish to thank 
SLAC for its support and kind hospitality. 
This work is supported by
DOE
and NSF (USA),
NSERC (Canada),
CEA and
CNRS-IN2P3
(France),
BMBF and DFG
(Germany),
INFN (Italy),
FOM (The Netherlands),
NFR (Norway),
MES (Russia),
MEC (Spain), and
STFC (United Kingdom). 
Individuals have received support from the
Marie Curie EIF (European Union) and
the A.~P.~Sloan Foundation.

%% file: prl.bbl
\begin{thebibliography}{99}
\bibitem{cc}
  Inclusion of the charge conjugate mode is implied.
\end{thebibliography}
